\documentclass{PoS}

\title{RXTE spectra of the Galactic microquasar GRO J1655--40 during
the 2005 outburst}

\ShortTitle{RXTE Spectra of GRO J1655--40}

\author{\speaker{Koji Saito}, Kazutaka Yamaoka, Mizuki Fukuyama,
Takehiro G. Miyakawa and Atsumasa Yoshida
	\\ Department of Physics and Mathematics, Aoyama Gakuin
University, Sagamihara, Japan\\ E-mail:
\email{ksaito@phys.aoyama.ac.jp}, \email{yamaoka@phys.aoyama.ac.jp},
\email{mfukuyama@phys.aoyama.ac.jp},
\email{tmiyakawa@phys.aoyama.ac.jp},
\email{ayoshida@phys.aoyama.ac.jp}}

\author{Jeroen Homan\\
        MIT Kavli Institute for Astrophysics and Space Reseach,
	Cambridge, USA\\ E-mail: \email{jeroen@space.mit.edu}}

\abstract{We report on the results of a detailed spectral analysis of
389 \textit{RXTE} observations of the Galactic microquasar GRO
J1655--40, performed during its 2005 outburst. The maximum luminosity
reached during this outburst was 1.4 times higher than in the
previous (1996--1997) outburst. However, the spectral behavior during
the two outbursts was very similar. In particular, $L_{\rm disk}$ was
proportional to $T_{\rm in}^4$ up to the same critical luminosity and
in both outbursts there were periods during which the energy spectra
were very soft, but could not be fit with standard disk models.}

\FullConference{VI Microquasar Workshop: Microquasars and Beyond\\
                 September 18-22, 2006\\
                 Como, Italy}

\begin{document}

\section{Introduction and Observations}

The black-hole transient GRO J1655--40 is a well known Galactic
superluminal jet source \cite{1}\cite{2}. After its 1996--1997 outburst 
it remained in quiescence for more than 7 years, until \textit{RXTE}/ PCA
detected a rise in the X-ray flux on 2005 February 17 \cite{3}.
Figure 1 shows {\it RXTE}/ASM light curves of the 1996--1997
and 2005 outbursts. The 2005 outburst lasted for 8 months and its
maximum luminosity was about a factor of 1.4 higher than in the
1996--1997 outburst (which had a duration of 16 months).

We have analyzed 389 \textit{RXTE}/PCA and HEXTE observations of GRO
J1655--40, which were carried out between 2005 March 7 and 2005
September 17, and 80 observations from the period between 1996 May 9
and 1997 September 11.  Spectra were extracted using FTOOLS v5.3.1.
PCA spectra were produced using 'Standard 2 mode' data from PCU2. We
corrected not only for PCA dead-time but also for
pile-up\footnote{From J. A. Tomsick and P. Kaaret (1998):
http://astrophysics.gsfc.nasa.gov/xrays/programs/rxte/pca/.} and a
2\% systematic error was added. We found that pile-up effects could
be as strong as  $\sim$10\%, but fitting parameters did not change
significantly even for such high values. HEXTE spectra were produced
from 'Archive mode' data of cluster A (0--3) and cluster B (0,1,3). 
The PCA data were fitted between 3 and 20 keV and the HEXTE data 
between 17 and 240 keV (or lower energies, in case of low source
counts).

\vspace{.4cm}
\begin{figure}[htbp] 
\begin{center}\hspace{-.2cm}
\includegraphics[width=15.3cm,keepaspectratio]{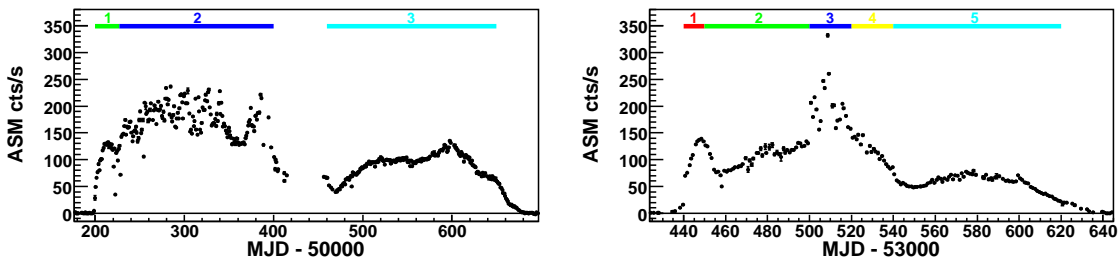}
\end{center} \caption{2-10 keV {\it RXTE}/ASM one-day averaged light
curves for GRO J1655--40. In the 2005 outburst (right panel), the
source returned to the hard state in about half the time compared to
the 1996--1997 outburst. The colors of the bars near the top of the
panels are explained in more detail in Section 2.}\label{1}
\end{figure}

\begin{figure}[t] \begin{center}
\includegraphics[width=15.2cm]{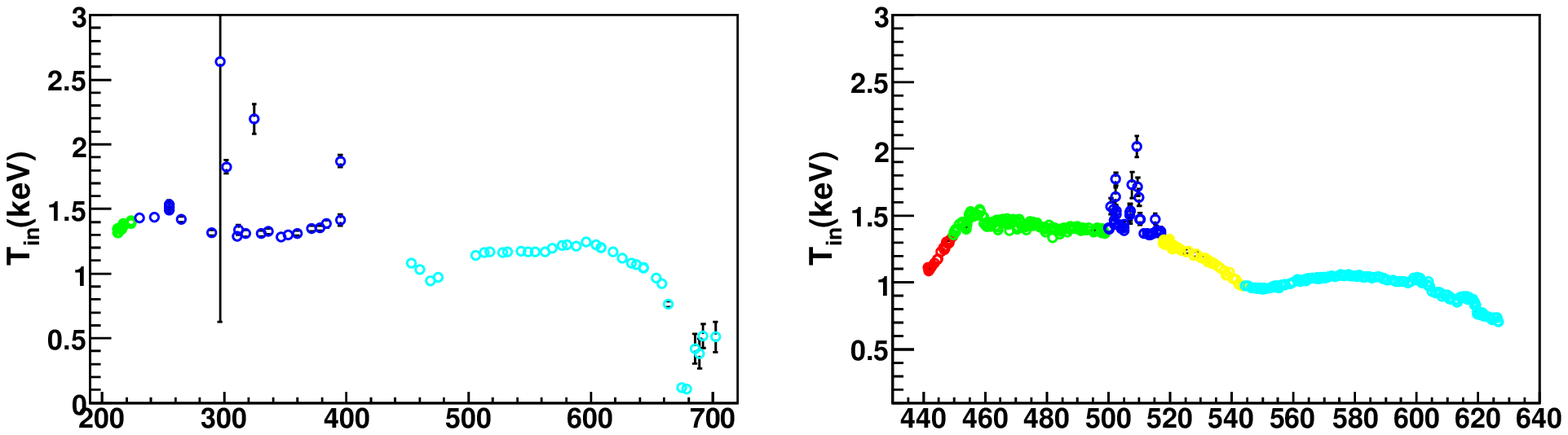}
\includegraphics[width=15.2cm]{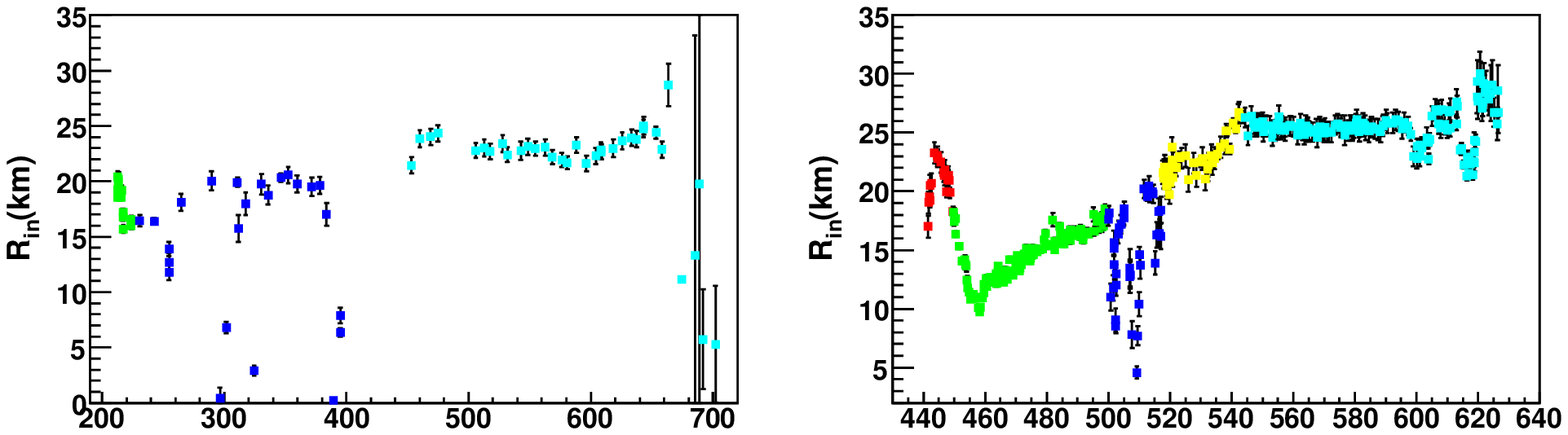}
\includegraphics[width=15.2cm]{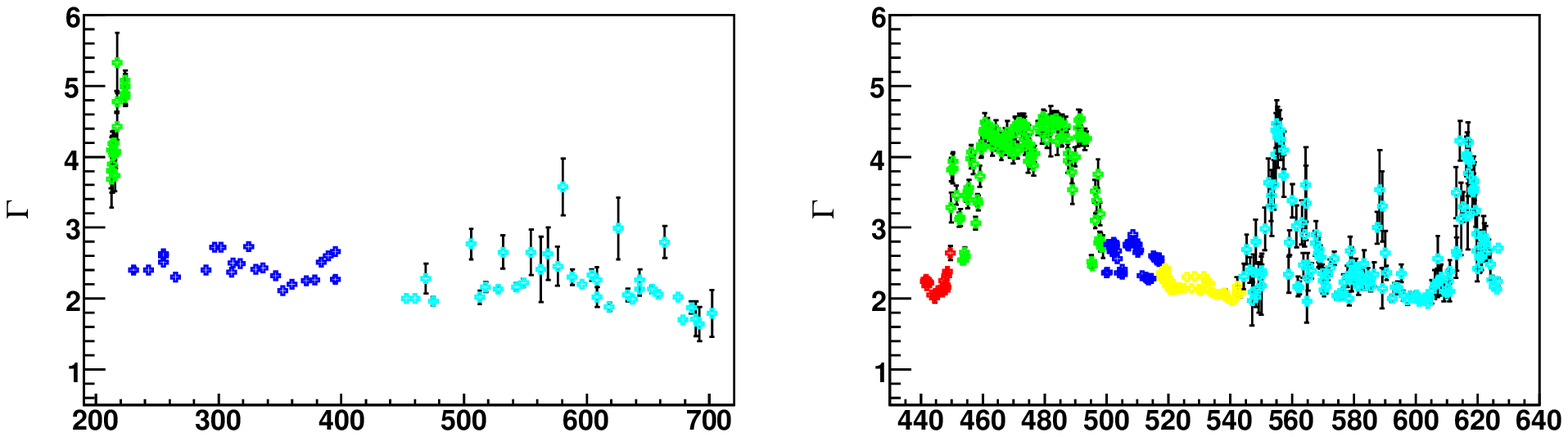}
\includegraphics[width=15.2cm]{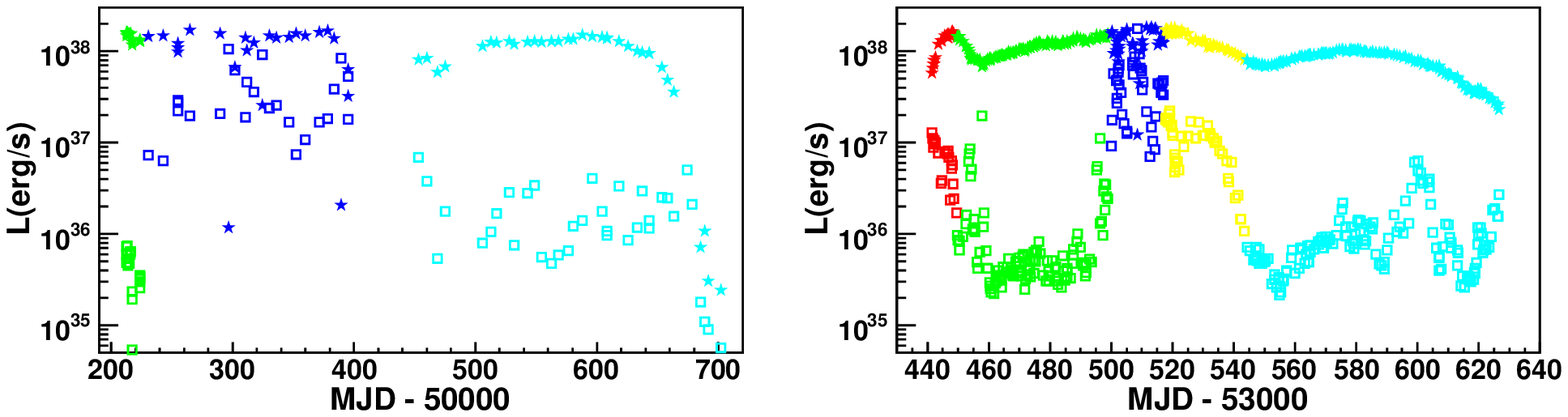} \end{center}
\caption{Time history of the spectral parameters of GRO J1655--40
during the 1996--1997 (left) and 2005 (right) outbursts. From top to
bottom: (1) the inner disk temperature $T_{\rm in}$, (2) the inner
disk radius $R_{\rm in}$, (3) the photon index of the power-law
component, $\Gamma$, and (4) the disk luminosity $L_{\rm disk}$
(filled star) and the luminosity of 3-100 keV power-law component
$L_{\rm hard}$ (open square). Epochs were color-coded: \textit{red}
(Epoch 1), \textit{green} (Epoch 2), \textit{blue} (Epoch 3),
\textit{yellow} (Epoch 4), \textit{aqua} (Epoch 5) in 2005 and
\textit{green} (Epoch 1), \textit{blue} (Epoch 2), \textit{aqua}
(Epoch 3) in 1996--1997.}\label{2} \end{figure}

\section{Results}


We applied a simple phenomenological spectral model consisting of a
multi-color disk black body and a power-law component. Following
\cite{4} we added an absorption line at 6.8 keV and absorption edges
at 7.7 keV, 8.8 keV, 9.3 keV and 10.8 keV to our model. The hydrogen
column density ($N_{\rm H}$) was fixed to 7.1$\times 10^{21}$ atoms
cm$^{-2}$. We were mainly interested in the evolution of the disk
parameters and since no disk component could be detected in the hard
state of GRO J1655--40, observations from that state were excluded
from further consideration. Figure 2 shows the time evolution
of the spectral parameters during both outbursts and Figure 3
shows typical PCA and HEXTE energy spectra of GRO J1655--40 for the
2005 outburst. 

The values of $R_{\rm in}$ were corrected by $ R_{\rm in} =
\kappa^2\xi r_{\rm in} $ \cite{5}, where $\kappa$ is a spectral
hardening factor \cite{6},  $\xi$ is a correction factor for the inner
boundary condition \cite{7}, and $r_{\rm in}$ is the apparent inner
disk radius. Although the value of $\kappa$ depends on a luminosity,
we followed \cite{5} and corrected the values of $R_{\rm in}$ with
$\kappa=1.7$ and $\xi=0.41$ as an approximation. The values of
$R_{\rm in}$ and $L_{\rm disk}$ are calculated for D = 3.2
kpc \cite{2} and an inclination of 70$^\circ$ \cite{8}. If D $\leq$ 1.7
kpc \cite{9}, $R_{\rm in}$ and $L_{\rm disk}$ decrease by a factor of
$\sim$53\% and $\sim$30\%, respectively. When $\Gamma \geq 3$ and 
the power-law component could not be well constrained (i.e. when it
was weak), $\Gamma$ was fixed to 2.1, which was nominal soft-state
value.

Based on the spectral fit results, the observations of the 2005
outburst could be divided into 5 Epochs (this excludes the hard
state). Epoch 1: the hard component was strong and the disk component
was underestimated ($R_{\rm in}$ was smaller than the value in
Epoch 5,  which is discussed below). Epoch 2: the spectra were very
soft and the source was rarely detected with HEXTE. $R_{\rm in}$
was small  ($\sim$ 10--18 km) and $T_{\rm in}$ was high ($\sim$
1.4--1.5 keV).  During the transitions from Epoch 1 to Epoch 2 and from
Epoch 2 to Epoch 3 the hardness fluctuated, so sharp boundaries between
these Epochs were difficult to define. Epoch 3: the luminosity
reached its maximum of the 2005 outburst and power-law component
increased dramatically. $R_{\rm in}$  and $T_{\rm in}$ exhibited
large fluctuations. Epoch 4: the hard component  was strong and
similar to Epoch 1. Epoch 5: $R_{\rm in}$ was fairly  constant at
$\sim$26km and $T_{\rm in}$ changed in accordance with the changes of
$L_{\rm disk}$. Although the hardness was not constant and  there
were days that $\Gamma$ was large (but not well constrained), the 
parameters of the disk component changed little and fitting was still
acceptable when $\Gamma$ was fixed at 2.1. Since the disk component
dominated the X-ray spectrum in Epoch 5, we consider the fitted
disk parameters to be more reliable than in the other Epochs.

The 1996--1997 outburst could be  roughly classified into 3 Epochs.
Epoch 1, 2 and 3 in the 1996--1997 outburst were similar to Epoch 2, 3
and 5 in the 2005 outburst, respectively. GRO J1655--40 therefore
passed  through similar states in the roughly the same sequence in
both outbursts.  However, there was no period in the 1996--1997
outburst corresponding to Epoch 1 in the 2005 outburst, probably
because the initial phase of the 1996--1997 outburst was not well
covered by pointed {\it RXTE} observations. In addition, it is not 
clear whether the 1996--1997 outburst has a period similar to Epoch 4 
in the 2005 outburst because the object went into the solar exclusion 
zone after Epoch 2 in the 1996--1997 outburst \cite{10}.

Although we tried to add emission lines and/or absorption
edges \cite{10}\cite{11}\cite{12}, our fits for Epoch 2 in the 2005
outburst and Epoch 1 in the 1996--1997 outburst were poor
($\chi^2/d.o.f \sim$ 2-4). As can be seen from Figure 4, when
the power-law component became strong, the disk component was
underestimated and $R_{\rm in}$ decreases. However, the distribution
of Epoch 2 in 2005 and Epoch 1 in 1996--1997 was quite different from
other Epochs in the sense that they did not fall on the main branch
traced out by the other Epochs. Perhaps the state of the accretion
disk was different in these two Epochs and other disk models need to
be applied.



Finally, Figure 5 shows the correlation between $T_{\rm in}$
and $L_{\rm disk}/L_{\rm E}$. $L_{\rm E}$ was calculated for the mass
of the  compact object $M =7M_\odot$ \cite{8}.  By comparing the 2005
outburst with the 1996--1997 outburst, we find that both outbursts
traced out very similar tracks and have nearly the same critical
luminosity ($\sim0.2L_{\rm E}$) at which the source starts to leave
the $L_{\rm disk} \propto T_{\rm in}^4$ relation [5]. This result 
indicates that there is likely a fixed physical process that causes 
the departure from the $L_{\rm disk} \propto T_{\rm in}^4$ relation 
in both outbursts.



\vspace{1cm} {\it Acknowledgment}: We would like to thank J. A.
Tomsick for providing his PCA pile-up correction program. This
report has made use of data obtained through the High Energy
Astrophysics Science Archive Research Center on-line service,
provided by NASA/Goddard Space Flight Center.

\begin{figure}[htbp]\hspace{-.5cm}
 \begin{minipage}{0.48\hsize}
  \begin{center}
    \includegraphics[width=5.2cm,angle=270,keepaspectratio]{spectra.eps}
  \end{center}
  \caption{Typical energy spectra of GRO J1655--40 in the 2005 outburst. The filled circles show PCA spectra and open squares show HEXTE spectra. Each color of the spectrum is correspond to the color used in Figure 2.}\label{fig:one}
 \end{minipage}\hspace{.7cm}
 \begin{minipage}{0.48\hsize}\vspace{-.7cm}
  \begin{center}
   \includegraphics[width=7.5cm,keepaspectratio]{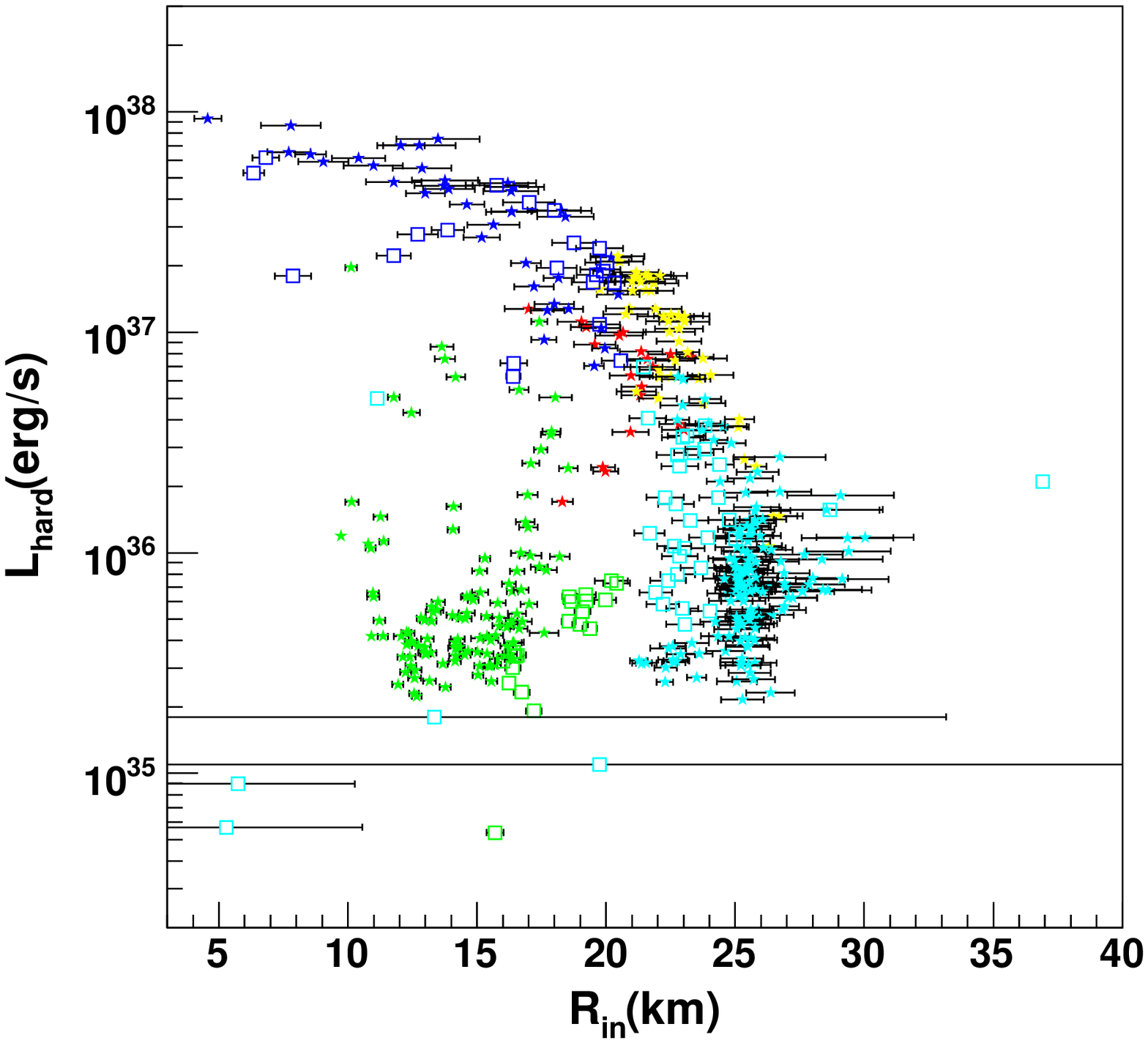}
  \end{center}
  \caption{Plot of $L_{\rm hard}$ vs $R_{\rm in}$. The data of the 1996--1997 outburst are plotted by open squares and the 2005 outburst are plotted by filled stars. The colors of the Epoch are same as Figure 2.}\label{fig:two}
 \end{minipage}
\end{figure}

\begin{figure}
 \begin{center}
  \includegraphics[width=11cm,keepaspectratio]{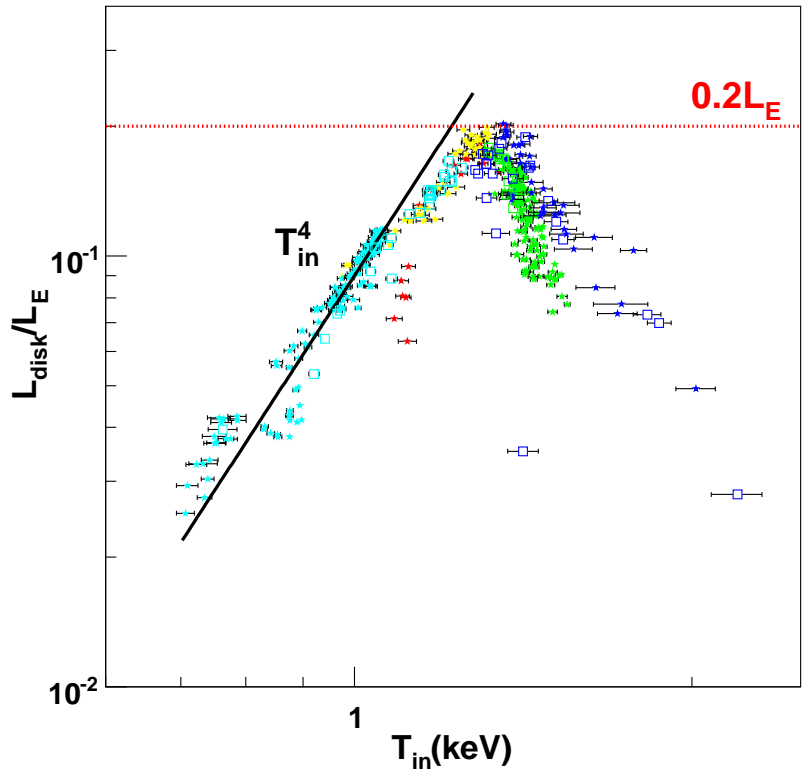}
 \end{center}
 \caption{Correlation between $L_{\rm disk}/L_{\rm E}$ and $T_{\rm in}$. The plotted symbols are same as Figure 4 and the colors of the Epoch are same as Figure 2. }\label{5}
\end{figure}

\end{document}